\documentclass[journal]{IEEEtran}
\usepackage{mathtools,booktabs,amsmath, cite, graphicx, array, stfloats, url, inputenc, mathrsfs,color,verbatim,multirow,algorithm,setspace,graphicx,subfigure}
\usepackage[justification=centering]{caption}
\usepackage{algorithm}
\usepackage{algorithmic}
\usepackage{multirow} 
\usepackage{amsmath} 
\usepackage{verbatim}
\usepackage{xcolor}

\ifCLASSINFOpdf
\else
\fi

\begin{document}

\title{When Six Degrees of Separation Meets Online Social Networks: How Low Can the Degree Be?}

\author{Zhimin~Zhang, 
        Fang Zhou,
        and Huansheng Ning 
\thanks{Z. Zhang and F. Zhou are with the School of Computer and Communication Engineering, University of Science and Technology Beijing, Beijing 100083, China.}
\thanks{H. Ning is with the School of Computer and Communication Engineering, University of Science and Technology Beijing, Beijing 100083, China, and also with the Beijing Engineering Research Center for Cyberspace Data Analysis and Applications, Beijing 100083, China.}
\thanks{\emph{(Corresponding authors: zhoufang@ies.ustb.edu.cn)}}
}

\maketitle

\begin{abstract}
  The proposal of Six Degrees of Separation makes people realize that the world is not as big as we imagined. 
  Even if the world's population now exceeds 7 billion, two strangers can still get in touch through a limited intermediary. 
  When online social networks have taken the world by storm, can people connect through shorter distances?  
  This issue is worth thinking about. This paper describes Six Degrees of Separation and the limitations of this theory in practical applications. 
  Combined with online social networks, the paper analyzes the development and change of the degrees of connection between people. 
  Finally, the paper considers the actual coverage of online social networks and other issues, and rationalizes the possible degrees of the future Six Degrees of Separation.
\end{abstract}


\IEEEpeerreviewmaketitle

\section{Introduction}
How big is the world? How many intermediaries do we need to reach a stranger? In 1929, Hungarian writer Frigyes Karinthy first proposed the Six Degrees of Separation in his book ``Chains'' \cite{karinthy1929chains}. He believed that all people on the planet could be connected to any other person through an acquaintance chain within six. In 1967, American sociologist Stanley Milgram designed a set of experiments to verify the authenticity of this theory \cite{guiot1976modification}. The investigation took the correct delivery of packages as the ultimate goal. Among all the successful results, it showed an average degree of 6.2, which verified the effectiveness of Six Degrees of Separation.

Six Degrees of Separation can explore the social distance between people and be used to analyze the cooperation between researchers under the same funding \cite{kardes2014complex,kardes2012six}, the relationship between tutors and students in specific fields \cite{aylward2012six}, and even the degree of collaboration between two universities in different disciplines \cite{leonesi2015mystery}. 
Six Degrees of Separation can be derived by using mathematical formulas. Assuming that the number of relatives and friends is $k$, the average degree of separation is $d$, and the maximum number of people that can be contacted is $N$. These three indicators satisfy the following relationship:
\begin{equation}
  N = k^d
\end{equation}

Assuming that the current world population is 8 billion, ideally, everyone needs to have more than 44.72 (that is, 45) relatives and friends to achieve Six Degrees of Separation. According to the description of Rule Of 150 (Dunbar's Number), human ability allows one to have a stable social relationship of about 150 people, so it is feasible to analyze the theory of Six Degrees of Separation from a theoretical perspective.

When Milgram was designing experiments, large-scale integrated circuits were just born, and computers began to enter a stage of rapid development. Computers at this time had a vast appearance and were generally used for scientific calculations. Milgram did not realize that this technology would bring earth-shaking changes to interpersonal communication and social connections, and even the description of the ``global village" appeared. The world seems to have become smaller under the influence of online social networks. People can use rich social network resources and powerful search engines to contact unfamiliar target candidates more easily.

Is Six Degrees of Separation still applicable in online social networks? How much impact will Six Degrees of Separation have? Does it shorten the social distance between people as people imagine? To explore these issues, researchers need to combine social platforms and data to recalculate the social distance between people.

This paper reviews a series of work on social distance evaluation in online social networks. It describes the development and changes of Six Degrees of Separation under different social network platforms and scales. The paper summarizes the lack of considerations in the Six Degrees of Separation experimental design and the level to which these factors affect degree in online social networks. In addition, the paper combines the development trend of online social networks and essential factors to speculate on the possible limit value of social distance in the future. The main contributions of this paper include:
\begin{enumerate}
  \item Summarizing the limiting factors of Six Degrees of Separation in practical applications and whether these factors impact the distance of interpersonal relationships in online social networks.
  \item Reviewing the Six Degrees of Separation changes with social platforms and user scales in the development of online social networks.
  \item Speculating how the average degree of interpersonal communication will drop in the future according to the development trend of online social networks.
\end{enumerate}

The other sections of this paper are arranged as follows. Section II describes the limits of Six Degrees of Separation and their influence on online social network analysis. Section III describes the performance of Six Degree of Separation in online social networks. Section IV speculates how the degree of interpersonal communication will change under the further expansion of online social network influence. Finally, the conclusion of this paper is given in Section V.

\section{Limitations of the Six Degrees of Separation and their influence on online social networks}

The proposal and verification of Six Degrees of Separation are shown intuitively that the world is much smaller than imagined. However, the experimental results are obtained in the ideal state, and in the actual social network, some factors will directly affect the average distance of interpersonal communication. This section summarizes these factors and analyzes the impact of these factors on social length in online social networks.

\subsection{Success rate and industry coverage issue}
In Milgram's experiment, a total of more than 300 volunteers participated. Among them, the number of packages delivered is about 60, and the success rate of the investigation is less than 25\%. The average social distance in the Six Degrees of Separation is calculated only in the successful results, and the unsuccessful experimental results are directly ignored. In addition, the target of delivery is a stockbroker, and the choice of industry coverage is minimal. Since people engaged in this industry often have a relatively large communication range, it may reduce the average social distance between interpersonal interactions. If changing the target to another industry, whether it will affect the average degree of separation has not been effectively verified.

This limitation can be weakened in social distance analysis based on online social networks. On the one hand, the scale of users in social network platforms is relatively large. A large amount of user data are readily available, which facilitates the extraction of complete users' social relationships, so the scale of the experiment can be effectively improved. On the other hand, due to the low threshold for user registration in online social networks, the user industry has wide coverage. It is not limited to a specific professional environment conducive to providing a more convincing average degree of separation.

\subsection{Social cost issue}
Six Degrees of Separation is a hypothesis put forward on the premise of undifferentiated interpersonal relationships. However, in actual social relationships, social cost is an essential factor restricting people's communication. Even if they are all relatives and friends, the strength and weaknesses of their relationship cannot be the same. Therefore, such social cost needs to be fully considered when calculating the degree of separation.

Social cost also exists in online social networks, even more evident than interpersonal communication in real life. 
For example, online conversations initiated to general friends may be ignored or even blacklisted due to concerns and other factors. 
Therefore, it is important for researchers to pay attention to this issue when evaluating the average degree of separation in online social networks. 
Ke \cite{ke2010social} considered this issue in their self-created instant messaging system and introduced a damping factor to express the social cost. 
When the damping factor increases, it means that the social cost between two people increases, and the possibility of effective social interaction decreases. 
Dara et al. \cite{DARAGHMI2014273} also used strong and weak connections to represent social costs when verifying the performance of Six Degrees of Separation in online social networks.

\subsection{Interpersonal network structure issue}
Six Degrees of Separation assumes that the interpersonal network is a tree structure, which is somewhat deviated from the actual interpersonal network structure. In real life, some interpersonal relationships have a loop structure, and even some people are in a closed-loop state. Therefore, when calculating the average degree, the actual structure of the interpersonal network needs to be considered.

N. Toyota \cite{toyota2008some} believed that the degree of clustering between nodes in the interpersonal network (i.e., clustering coefficient) and whether the clustering coefficient would affect the ability of information dissemination should be effectively considered. Through experimental verification, he believed that the clustering coefficient was not a critical element that affects information dissemination.

To meet the clustering requirements and reduce the average path length, Duncan J. Watts and Steven Strogatz proposed a graph structure with small-world network attributes, namely Watts-Strogatz \cite{watts1998collective}. This graph structure is between the regular and random graphs, taking into account the distance between relatives and friends in virtual social networks. In order to construct Watts-Strogatz, three parameters need to be determined: the number of nodes $N$, the number of nearest neighbors $K$ of a node, and the probability $p$ of randomized reconnection. The introduction of $p$ meets the requirements of virtual social networks, that is, individuals have a gathering relationship with their relatives and friends, but they may also have a long-distance from some relatives and friends due to geography and other factors, and $p$ is responsible for constructing these long distances. When $p=0$, Watts-Strogatz degenerates into a regular graph, and the average path length between nodes at this time is $d=N/2K$. As $p$ continues to increase, the randomness of the graph also continues to increase. When $p=1$, the graph is completely random, and the average path length between nodes at this time is $d=ln(N)/ln(K)$.

\begin{figure}[ht]
  \centering
  \includegraphics[scale=0.065]{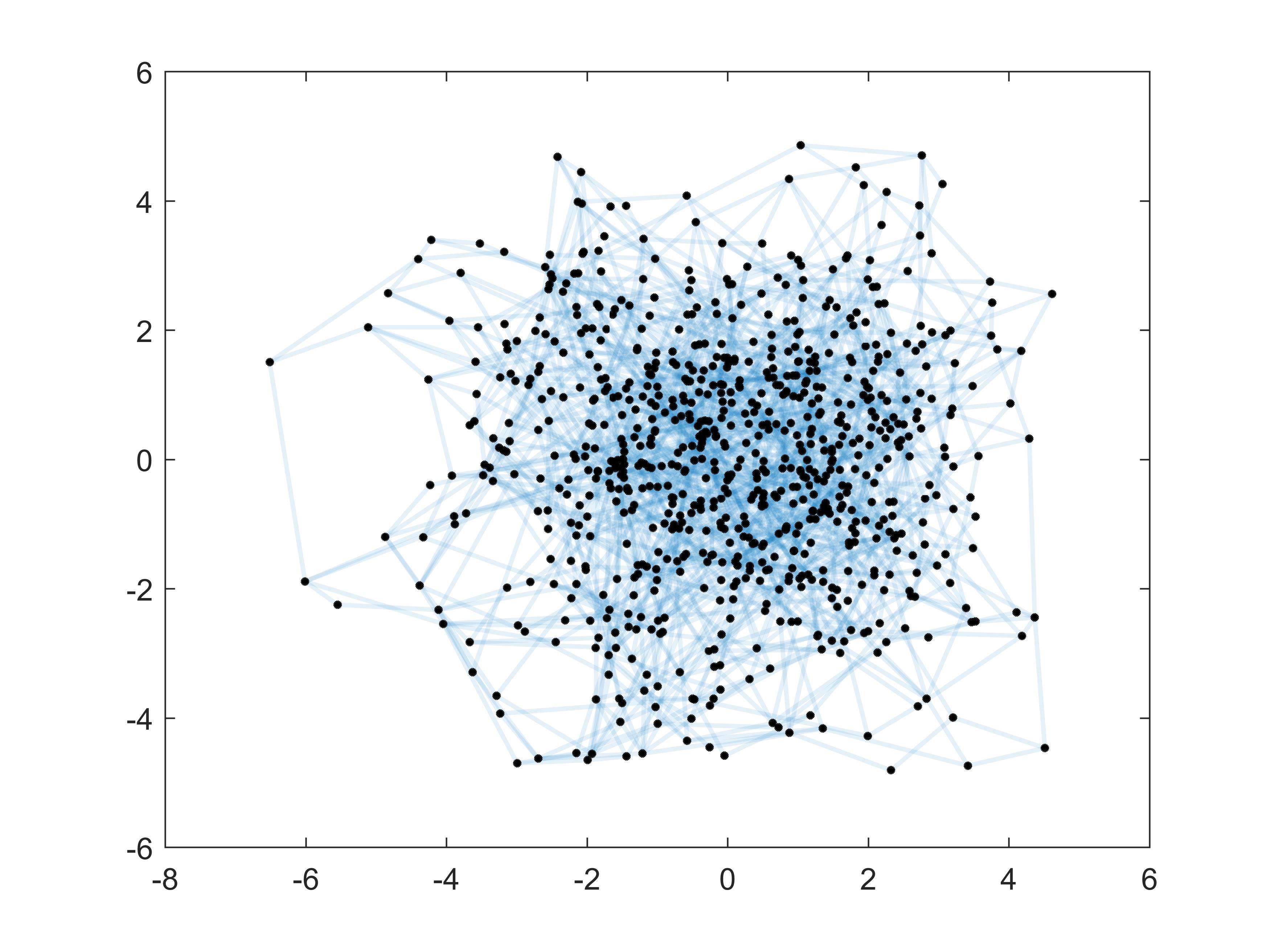}
  \caption{Watts Strogatz based on 800 people ($K=4, p=1$)}  \label{fig1}
  \end{figure}

Taking a small social network of 800 people as an example, based on Six Degrees of Separation, a person needs to know at least 3.04 (that is, 4) individuals to realize the theory. In Watts-Strogatz, when $p=1$, the average path length of the 800 people is 6.012, which is in line with the theoretical value. The Watts-Strogatz simulated in this state is shown in Fig. \ref{fig1}. Besides, according to the Six Degrees of Separation, under the assumption that there are 8 billion people globally, a person needs to know at least 43 relatives and friends to realize this theory. If $N$ is set to 8 billion and $K$ is set to 43, the average path length based on $p=1$ is 6.063, consistent with the experimental results of Six Degrees of Separation. However, this is the result obtained under the completely random situation of Watts-Strogatz, and does not reflect the actual online social network situation. Therefore, after fully considering the structure of the interpersonal network, the average degree of separation in online social networks may change.

\subsection{Social differentiation issue}
The verification of Six Degrees of Separation was carried out in an entirely idealized state. It did not consider the influence of external factors such as race, economy, and politics on the average degree of separation. Some countries and regions have more serious social differentiation, which may be a direct factor for the rise of degree. Therefore, Six Degrees of Separation may only be applicable in certain classes and environments.

These external factors do not seem to be improved in the online social network environment. For example, some social platforms have opened a membership mechanism to classify users according to their activity level, recharge level, and other factors, giving users of different levels different privileges, which will become one of the reasons that hinder users from socializing. In addition, due to differences in living environment, cognitive abilities, etc., different users will have different views on the same event or even wholly opposing views. Different public opinions will also affect users' social behaviors, which in turn affects the average degree of separation.

\section{Six Degrees of Separation influenced by online social networks}
The rise of online social networks has set off an upsurge of instant messaging, which seems to shorten the distance between humans and facilitate the interaction between people. Under the influence of this new social mode, how the degree of separation between interpersonal communication has changed has also inspired scholars' discussion and study.

At the same time, technologies such as data mining, community mining, and precise recommendation began to emerge in online social networks, providing convenient conditions for quickly matching similar users. The emergence of swarm intelligence methods such as ant colony optimization algorithm and particle swarm optimization algorithm has accelerated the search speed of user relevance and reduced the cost of communication between people. There are two recommendation methods in online social networks that can be used to expand user communication scope to reduce the degree of separation between users, that is, mutual friend recommendation and potential friend recommendation.

\subsection{Mutual friend recommendation}
According to the ternary closure theory, if two strangers in a social circle have common friends, the possibility of the two people establishing a connection will increase. The magnitude of this possibility is obviously related to the number of mutual friends. Through mutual friend recommendation, some users can skip an intermediate contact and directly establish connection with the target user. In this case, the degree of separation can be reduced from 1 to 0. Mutual friend recommendation has been applied in many mainstream social networks, such as Tencent QQ, TikTok, etc.

\subsection{Potential friend recommendation}
Some social networks have prominent community characteristics, such as Facebook, Twitter, Instagram, etc. In these platforms, potential friend recommendation has become a new measure to reduce the degree of separation. Calculating the similarity between users through various indicators such as user behavior, user attention content, user interest, user topics, and user interaction status, mining potential relationships, and making recommendations, can effectively promote user interaction and expand the scope of user communication.

Taking Facebook as an example, we investigated the performance of this social platform based on the average degree of separation in recent related work and summarized them in Table \ref{Changes in the degrees of separation of the same social platform}.

\renewcommand\arraystretch{1.3} 
\begin{table}[htbp]
  \centering
\caption{Changes in the degree of separation of the same social platform (take Facebook as an example)} \label{Changes in the degrees of separation of the same social platform}
\resizebox{\linewidth}{!}{
  \begin{tabular}{cccc}
  \toprule
  \toprule
  Year      & Platform user scale     & \begin{tabular}[c]{@{}c@{}}Average degrees \\ of separation\end{tabular} & Reference \\ \midrule
  2008      & 100 million                    & 5.28                                                                     & \cite{6425590}          \\
  2010      & 600 million                   & 5.06                                                                     & \cite{6425590}          \\
  2011      & 721 million           & 4.74                                                                     & \cite{10.1145/2380718.2380723}          \\
  2012-2013 & 950 million           & 3.868                                                                    & \cite{DARAGHMI2014273}          \\
  2014      & 1 billion 300 million & 5.0                                                                      & \cite{7159171}          \\
  2016      & 1 billion 590 million & 3.57                                                                     & \cite{edunov2016three}          \\ \bottomrule \bottomrule
  \end{tabular}}
  \end{table}

  As can be seen from Table \ref{Changes in the degrees of separation of the same social platform}, as the years go on, the user scale of Facebook has been growing. The average degree of separation based on this platform fully conforms to the six degrees of separation and shows a declining state of degrees over time; that is, the cost of communicating with unfamiliar users is getting lower and lower.

  There are three reasons for the overall decline in Facebook's average degrees of separation:
  \begin{enumerate}
    \item The continuous increase in the number of registered users provides suitable conditions for expanding the scope of user communication and tapping potential user relationships.
    \item Facebook continues to introduce new social functions, facilitating the development of potential friend recommendations and other related work. For example, in 2019 alone, Facebook launched more than a dozen significant features such as the ``memorial" area, ``birthday story," Facebook news, Facebook dating, etc. They enhanced Facebook's community characteristics.
    \item Data mining, community mining, and the continuous update of recommendation algorithms have further improved precise and personalized recommendation.
  \end{enumerate}

  \renewcommand\arraystretch{1.3} 
  \begin{table}[htbp]
    \centering
    \caption{Changes in the degree of separation of the different social platforms} \label{Changes in the degrees of separation of the different social platforms}
    \resizebox{\linewidth}{!}{
    \begin{tabular}{ccccc}
    \toprule \toprule
    Year & Platform                                                                                            & \begin{tabular}[c]{@{}c@{}}Experiment \\ scale\end{tabular}      & \begin{tabular}[c]{@{}c@{}}Average degrees \\ of separation\end{tabular}                                                                                             & Reference \\ \midrule
    2006 & \begin{tabular}[c]{@{}c@{}}Microsoft\\ Messenger\end{tabular}                                       & 180 million           & 6.6                                                                                                                                                                  & \cite{whoriskey2008instant}          \\ \midrule
    2009 & ArnetMiner                                                                                          & 0.5million            & \begin{tabular}[c]{@{}c@{}}\textless{}3\\  (Regardless of the strength \\ of the relationship)\\ $\approx$6\\ (Considering the strength \\ of the relationship)\end{tabular} & \cite{zhang2009six}          \\ \midrule
    2009 & Tencent QQ                                                                                          & 783.4 million         & 3.12                                                                                                                                                                 & \cite{zhang2009six}          \\ \midrule
    2011 & Twitter                                                                                             & -                     & 3.43                                                                                                                                                                 & \cite{bakhshandeh2011degrees}          \\ \midrule
    2012 & \begin{tabular}[c]{@{}c@{}}Institutions funded \\ by the National\\ Science Foundation\end{tabular} & 279862                & \begin{tabular}[c]{@{}c@{}}3.07\\ 2.75\\ (No organization that only \\ cooperates once or twice)\end{tabular}                                                        & \cite{kardes2012six}          \\ \midrule
    2016 & Facebook                                                                                            & 1 billion 590 million & 3.57                                                                                                                                                                 &  \cite{edunov2016three}         \\ \bottomrule \bottomrule
    \end{tabular}}
    \end{table}

    Table \ref{Changes in the degrees of separation of the different social platforms} summarized the average degrees of separation of different social platforms at a certain point of time. Although the degrees of separation on each platform are not the same, they generally meet the Six Degrees of Separation. In addition, there are significant differences in the degrees of separation between social platforms, mainly for three reasons:
    \begin{enumerate}
      \item The scale of social platform users. The larger the user scale, the lower the theoretical average degree of separation. The user scale will directly determine the complexity of the relationship and the difficulty of communication between strangers.
      \item The degree of social content aggregation. Taking \emph{Institutions funded by the National Science Foundation} as an example, since the research subjects are only for institutions funded by the same fund, the user type and social purpose are single, which will reduce the average degree of separation between users.
      \item The strength of the relationship. When considering the strength of interpersonal relationships, even the same social platform will show differences in the degrees of separation. For example, in ArnetMiner (researcher academic search network), the average degree of separation is only $<$3 regardless of the strength of the relationship, and when the relationship strength is considered, the average degree of separation is $\approx$6, and the gap between the two is significant.
      In addition, according to the discussion results in Section II, the social cost issue should be effectively considered when calculating the degree of separation, which will be more in line with actual social network requirements.
    \end{enumerate}

    When calculating the degree of separation between users based on a specific social platform, the representation of social data and the calculation of separation need to be considered.
    For the term of social data, most studies use graphical structure.
    This directly represents whether people have contact in social networks, which is convenient for analyzing the degree of separation between users.
    Graph databases such as Neo4j can be used to store corresponding data \cite{DARAGHMI2014273}. 
    When the user's social network is too extensive, tools such as WebGraph can also be used for graph compression \cite{10.1145/2380718.2380723}.
    The purpose of social data representation is to facilitate the solution of the degree of separation. 
    Ke et al. \cite{ke2010social} analogized the path problem between individuals to the routing problem in a computer network and used a modified routing table to represent user relationship data.
    Bakhshandeh et al. \cite{bakhshandeh2011degrees} directly sent information requests to Twitter via the Internet, avoiding representation and storage of user relationship data.

    No matter what method is used to obtain, represent, and store social relationship data, the ultimate goal is to calculate the degree of separation.
    On the one hand, the degree of separation can be calculated based on the user's friends and the target objects of interest.
    On the other hand, the degree of separation can be derived based on the user's geographic location.
    It has been discovered that geographical location and social relations are closely related and play an essential role in forming the entire social network structure \cite{10.1145/2380718.2380723}.
    The study by Backstrom et al. \cite{backstrom2010find} found that users who are physically close to each other may interact more frequently on Facebook, so it is also feasible to use geographic location information to calculate the degree of separation.

    The mainstream degree of separation calculation methods include greedy algorithms and search algorithms. 
    Using these technologies is because the Six Degrees of Separation can be abstracted as the shortest path problem.
    Analogous to the routing algorithm, Ke et al. \cite{ke2010social} used the Bellman-Ford algorithm to find the shortest path between two users.
    Bakhshandeh et al. \cite{bakhshandeh2011degrees} compared the performance of the greedy algorithm based on maximum attention counts, the greedy algorithm based on geographic heuristics, and the greedy algorithm based on boundary search in calculating the degree of separation. 
    Their real-time costs cannot meet the requirements of online evaluation applications.
    In addition, they also compared the two-way search method based on breadth-first and the two-way search method based on probability. 
    They found that the average degree search time of the last method is less than 15 seconds, which could be used for online separation evaluation.
    Based on a heuristic search method, Lawrence et al. \cite{lawrence2015analysis} used the ant colony optimization technology to determine the distance between two persons in the graph.

    In addition, Edunov et al. \cite{edunov2016three} used the Flajolet-Martin algorithm to estimate the number of unique friends for each degree of separation based on probability in big data to evaluate the overall degree of separation of the Facebook friendship graph.
    Based on graph compression technology and diffusion calculation, Backstrom et al. \cite{10.1145/2380718.2380723} used the HyperANF probability algorithm to analyze the degree of separation.
    By calculating the degree of separation in all the above methods, it is found that the Six Degrees of Separation is fully applicable in online social networks.

    With the development of computer technologies and a series of optimization algorithms, the degrees of separation in most social platforms have been lower than 6. \emph{four degrees of separation} \cite{10.1145/2380718.2380723,DARAGHMI2014273}, \emph{three and a half degrees of separation} \cite{7159171}, and other lower degrees of separation theory are put forward.
    This also indirectly shows that the social distance and social costs of people are further reduced due to the development of online social platforms.

\section{Possible conjectures on the future ``degree'' in online social networks}

The influence of online social networks continues to expand under the impetus of many emerging technologies, and people's dependence on these social network platforms is becoming stronger. In the future, how will the degrees of separation change under the influence of online social networks? Whether this value maintains the current trend and continues to decline or has reached the limit or even rebounded, this requires a rational analysis of the development trend of online social networks in the future.

\subsection{Influencing factors of ``degree'' in the future}
In addition to the user scale, industry coverage, and other factors mentioned above, the following factors will also affect the degree of separation in online social networks.

\subsubsection{Cross-application factor}
The performance of Six Degrees of Separation in online social networks has been studied based on a single platform. What cannot be ignored is that social network platforms have begun to appear in a situation of social integration and are showing an ever-expanding trend. The purpose of social integration, on the one hand, is to further make up for the lack of functionality of a single social platform; on the other hand, it is the process of integrating various interpersonal relationships among users. The cross-application development of social networks can open up interpersonal relationships restricted by specific platforms and unearth the potential relationships and connections between users and others on different social platforms, which will become a possible factor to reduce the degree of separation further.

\subsubsection{Rich media factor}
With the development of computer technologies such as streaming media, Flash, and Java, social networking platforms have gradually improved interactive functions such as animation and sound. This will help improve the user experience and increase the user's stickiness to the social platform. In addition, a good user experience helps to establish a good reputation among the user groups, which will become an implicit promotion model for social network platforms, which is conducive to expanding the scale of users on the platform.

\subsubsection{Registration convenience factor}
It is becoming easier and faster to register users' accounts on social network platforms. It may be necessary to fill in detailed user personal information in the past registration process, including gender, birthday, and other information. These cumbersome steps may cause some users to terminate registration, resulting in the loss of potential users. In the current user registration process, social platforms can rely on their cooperative relationship to directly fill user registration information through user authorization. For example, in the mini-program of WeChat, users can directly register and log in to the mini-program by asking whether they are authorized or not to realize the click-to-use function. This convenient operation is conducive to expanding the user scale of social platforms and provides conditions for reducing the degree of separation.

\subsubsection{Privacy policy factor}
In recent years, users' personal information has been leaked frequently, which has severely affected users' personal lives and the reputation of social network platforms. To reduce the occurrence of such incidents and as far as possible to allow users to choose whether to disclose certain privacy rights, various online social network platforms and operating systems are tightening privacy policies. This approach ensures user privacy security and increases user choice. For example, on April 27, 2021, Apple released a new generation of the operating system, iOS 14.5. The release of this version means that Apple's new privacy policy has officially taken effect. In this privacy policy, the application needs to obtain the user's consent before it can track the online behavior data of the mobile phone. This strategy will help protect user privacy, make application tracking strategies transparent, and improve users' control over their privacy. However, this is undoubtedly a huge impact on online social networks such as Facebook that need to use user online behavior data for personalized recommendations and advertisement. When users choose not to disclose their online behavior, this will affect the recommendation's accuracy and affect the degree of separation.

Table \ref{Table: Influence} summarizes the factors that may affect the degree of separation of online social networks in the future, the extent to which these factors affect the degree of separation, and the correlation between the two. For example, the user scale of online social networks will affect the degree of separation, and the influence of this factor is extremely strong. There is a negative correlation between this factor and the degree of separation; in general, the larger the user scale, the lower the degree of separation.

\begin{table}[htbp]
  \centering
  \caption{The influence of factors in the development of online social networks on the degrees of separation} \label{Table: Influence}
  \resizebox{\linewidth}{!}{
  \begin{tabular}{ccc}
  \toprule
  \toprule
  Influence factor                                                                          & Influence ability & \begin{tabular}[c]{@{}c@{}}Correlation with\\  degrees of separation\end{tabular} \\ \midrule
  User scale                                                                                & Extremely strong  & -                                                                                 \\
  Privacy policy                                                                            & Extremely strong  & +                                                                                 \\
  Recommendation algorithms                                                                 & Strong            & -                                                                                 \\
  \begin{tabular}[c]{@{}c@{}}Social cost \\ (strong or weak user relationship)\end{tabular} & Strong            & +                                                                                 \\
  Social network function                                                                   & Strong            & -                                                                                 \\
  Social platform cross-application                                                         & Strong            & -                                                                                 \\
  \begin{tabular}[c]{@{}c@{}}Public opinion \\ (social differences)\end{tabular}            & Mid               & +                                                                                 \\
  Rich media                                                                                & Mid               & -                                                                                 \\
  Convenience of registration                                                               & Mid               & -                                                                                 \\
  \begin{tabular}[c]{@{}c@{}}Industry coverage \cite{DARAGHMI2014273} \\ (including rare occupations)\end{tabular} & Weak              & +                                                                                 \\
  Social network structure \cite{toyota2008some}                                                                & Weak              & -                                                                                 \\
  Celebrity effect \cite{DARAGHMI2014273}                                                                         & Weak              & -                                                                                 \\
  \bottomrule \bottomrule
  \end{tabular}}
  \end{table}

\subsection{Possible conjectures on the future ``degree''}
There is no doubt that Six Degrees of Separation has full applicability in online social networks. Because the interaction methods of online social networks are quite different from real interactions and traditional interactions (such as phone calls, faxes, etc.), this makes the degrees of online social networks show a further downward trend and has different performance in different social platforms. In the future social networks, what is the limit value of the degrees? This issue requires a reasonable guess based on the development trend of social networks.

The above 12 factors in Table \ref{Table: Influence} are comprehensively considered, and they are divided into four categories according to future development trends; namely, enhanced positive correlation, weakened positive correlation, enhanced negative correlation, and weakened negative correlation.

\subsubsection{Enhanced positive correlation}
To better protect users' data, continuous improvement and refinement of privacy policies is the mainstream trend in the future. This trend will limit the platforms to analyze users' online behaviors, so it is an essential factor for increasing degree. Due to the large number of participants, the great freedom of expression, public opinion presents a complex trend, which will cause more serious social differentiation and will also lead to a trend of increasing degree. In addition, with the further growth of the user scale in online social networks in the future, the user industry coverage will be broader and will include more rare occupations. The increase of these rare occupations will result in an increase in degree, but the effect of this factor is weak and will not become the mainstream element of degree's development.

\subsubsection{Weakened positive correlation}
Although there are costs in communication between people, the costs show a decreasing trend with the development of online social technologies. Reducing the costs is conducive to building interactive chains between strangers so that the degree of separation is further reduced.

\subsubsection{Enhanced negative correlation}
The further improvement of registration convenience will stimulate more users to enter a specific social network, which will promote the expansion of user scale and present a more complex network social structure. In addition, social platform operators will continue to optimize their recommendation algorithms, expand the functions of social platforms, enhance cross-application capabilities, and improve rich media deployment to increase user stickiness. These factors will play a key role in reducing the degree of separation.

\subsubsection{Weakened negative correlation}
Diversified, virtualized, and anonymized communication methods enable ordinary people to become well-known, making the original celebrity effect composed of a small part of the leading individuals show a weakening trend. When the celebrity effect is weakened, the network structure that relies on celebrities will be reduced, which will increase the degree. Still, the influence of this factor is limited and will not affect the overall development trend of the degree.

By dividing and analyzing the development trend of these factors, combined with existing research results, it is a possible reasonable conjecture that the degrees in online social networks will be further reduced in the future. For comprehensive social platforms such as Facebook, Twitter, and Tencent QQ, the limit value of degree may be 3. For professional social platforms such as ArnetMiner and LinkedIn, the possible limit value of degree will not be lower than 2.5. The difference between the two types of platforms in the degrees of separation is that although the comprehensive social platforms have a large number of users, the relatively scattered social content will further limit the reduction of the degree. For professional social platforms, the social purpose is apparent. Although the user scale may be small, it may still show lower degree of separation.

\section{Conclusion}
This paper reviewed the performance of Six Degrees of Separation in online social networks. First, the paper described the Six Degrees of Separation and pointed out the limitations of the theory in practical applications. Subsequently, the paper summarized the performances of Six Degrees of Separation on different online social networks and gave reasonable explanations for the changes in degrees based on these performances. In addition, the paper summarized the factors that may affect the development of the degree in the future and reasonably speculated the limit value of the degree in the future. As online social networks continue to develop in-depth, the overall development trend of degree will continue to show a downward trend. The rate of decline will also continue to slow down due to bottlenecks such as technologies.

\ifCLASSOPTIONcaptionsoff
  \newpage
\fi



\bibliography{ref}
\bibliographystyle{ieeetr}

  \begin{IEEEbiographynophoto}{Zhimin Zhang}
    received the B.E. degree from Hebei University of Architecture, China, in 2019.
    He is currently working toward the Ph.D. degree from the School of Computer and Communication Engineering, University of Science and Technology Beijing, China.
    His research interests include Internet of Things and Artificial Intelligence.
    \end{IEEEbiographynophoto}

    \begin{IEEEbiographynophoto}{Fang Zhou}
    received the B.Sc, M.Sc and Ph.D degree in computer science from the University of Science and Technology Beijing, China, in 1995, 2002 and 2012. 
    From 2015 to 2016, she was a Visiting Researcher with the Department of Computer and Information Sciences, Temple University, USA. 
    She is currently an Associate Professor with the Department of Computer Science and Technology, University of Science and Technology Beijing. Her research interests include machine learning, information retrieval and information safty.
     \end{IEEEbiographynophoto}

\begin{IEEEbiographynophoto}{Huansheng Ning}
  [SM'13] received his B.S. degree from Anhui University in 1996 and his Ph.D. degree from Beihang University in 2001.
  He is currently a Professor and Vice Dean with the School of Computer and Communication Engineering, University of Science and Technology Beijing, China, and the founder and principal at Cybermatics and Cyberspace International Science and Technology Cooperation Base.
  He has authored 6 books and over 180 papers in journals and at international conferences/workshops. He has been the Associate Editor of IEEE Systems Journal, the associate editor (2014-2018), area editor (2020-) and the Steering Committee Member of IEEE Internet of Things Journal (2018-).
  He is the host of the 2013 IEEE Cybermatics Congress and 2015 IEEE Smart World Congress. His awards include the IEEE Computer Society Meritorious Service Award and the IEEE Computer Society Golden Core Member Award. His current research interests include Internet of Things, Cyber Physical Social Systems, Cyberspace Data and Intelligence. In 2018, he was elected as IET Fellow.
  \end{IEEEbiographynophoto}

\end{document}